\documentclass[a4paper]{jpconf}
\usepackage{graphicx}
\include{newcommands}

\begin{document}

\title{Measurement of the ${}^{92,93,94,100}$Mo($\gamma$,n) reactions by Coulomb Dissociation}

\author{
K~G\"{o}bel${}^{1}$, P~Adrich${}^{2}$, S~Altstadt${}^{1}$, H~Alvarez-Pol${}^{3}$, F~Aksouh${}^{2}$, T~Aumann${}^{2}$, M~Babilon${}^{4}$, K-H~Behr${}^{2}$, J~Benlliure${}^{3}$, T~Berg${}^{5}$, M~B\"{o}hmer${}^{6}$, K~Boretzky${}^{2}$, A~Br\"{u}nle${}^{2}$, R~Beyer${}^{7}$, E~Casarejos${}^{8}$, M~Chartier${}^{9}$, D~Cortina-Gil${}^{2}$, A~Chatillon${}^{3}$, U~Datta~Pramanik${}^{10}$, L~Deveaux${}^{11}$, M~Elvers${}^{4,12}$, T~W~Elze${}^{1}$, H~Emling${}^{2}$, M~Erhard${}^{7}$, O~Ershova${}^{1,2}$, B~Fernandez-Dominguez${}^{9}$, H~Geissel${}^{2}$, M~G\'orska${}^{2}$, T~Heftrich${}^{1}$, M~Heil${}^{2}$, M~Hellstroem${}^{2}$, G~Ickert${}^{2}$, H~Johansson${}^{2,13}$, A~R~Junghans${}^{7}$, F~K\"{a}ppeler${}^{14}$, O~Kiselev${}^{5}$, A~Klimkiewicz${}^{2}$, J~V~Kratz${}^{5}$, R~Kulessa${}^{15}$, N~Kurz${}^{2}$, M~Labiche${}^{16}$, C~Langer${}^{1,2}$, T~Le~Bleis${}^{2,17}$, R~Lemmon${}^{18}$, K~Lindenberg${}^{4}$, Y~A~Litvinov${}^{2}$, P~Maierbeck${}^{6}$, A~Movsesyan${}^{2,4}$, S~M\"{u}ller${}^{4}$, T~Nilsson${}^{13}$, C~Nociforo${}^{2}$, N~Paar${}^{19}$, R~Palit${}^{2}$, S~Paschalis${}^{9}$, R~Plag${}^{1,2}$, W~Prokopowicz${}^{2}$, R~Reifarth${}^{1,2}$, D~M~Rossi${}^{2,5}$, L~Schnorrenberger${}^{4}$, H~Simon${}^{2}$, K~Sonnabend${}^{1}$, K~S\"{u}mmerer${}^{2}$, G~Sur\'owka${}^{15}$, D~Vretenar${}^{19}$, A~Wagner${}^{7}$, S~Walter${}^{14}$, W~Walu\'s${}^{15}$, F~Wamers${}^{2}$, H~Weick${}^{2}$, M~Weigand${}^{1}$, N~Winckler${}^{14}$, M~Winkler${}^{2}$ and A~Zilges${}^{4,12}$}

\address{${}^{1}$ Goethe-Universit\"{a}t Frankfurt a. M., Germany}
\address{${}^{2}$ GSI Helmholtzzentrum f\"{u}r Schwerionenforschung, Darmstadt, Germany}
\address{${}^{3}$ Universidad de Santiago de Compostela, Spain}
\address{${}^{4}$ Technische Universit\"{a}t Darmstadt, Germany}
\address{${}^{5}$ Johannes Gutenberg-Universit\"{a}t Mainz, Germany}
\address{${}^{6}$ Technische Universit\"{a}t M\"{u}nchen, Germany}
\address{${}^{7}$ Helmholtz-Zentrum Dresden-Rossendorf, Germany}
\address{${}^{8}$ Universidade de Vigo, Spain}
\address{${}^{9}$ University of Liverpool, United Kingdom}
\address{${}^{10}$ SINP Kolkata, India}
\address{${}^{11}$ Universit\'{e} de Paris Sud, Orsay, France}
\address{${}^{12}$ Institut f\"{u}r Kernphysik, Universit\"{a}t zu K\"{o}ln, Germany}
\address{${}^{13}$ Chalmers University of Technology, G\"{o}teborg, Sweden}
\address{${}^{14}$ FZ Karlsruhe, Germany}
\address{${}^{15}$ Jagellonian University Krakow, Poland}
\address{${}^{16}$ University of Paisley, United Kingdom}
\address{${}^{17}$ University of Strasbourg, France}
\address{${}^{18}$ CCLRC Daresbury Laboratory, United Kingdom}
\address{${}^{19}$ University of Zagreb, Croatia}

\ead{goebel@physik.uni-frankfurt.de}

\newpage

\begin{abstract}
The Coulomb Dissociation (CD) cross sections of the stable isotopes ${}^{92,94,100}$Mo and of the unstable isotope ${}^{93}$Mo were measured at the LAND/R${}^{3}$B setup at GSI Helmholtzzentrum f\"{u}r Schwerionenforschung in Darmstadt, Germany. Experimental data on these isotopes may help to explain the problem of the underproduction of $^{92,94}$Mo and $^{96,98}$Ru in the models of p-process nucleosynthesis. The CD cross sections obtained for the stable Mo isotopes are in good agreement with experiments performed with real photons, thus validating the method of Coulomb Dissociation. The result for the reaction $^{93}$Mo($\gamma$,n) is especially important since the corresponding cross section has not been measured before. A preliminary integral Coulomb Dissociation cross section of the $^{94}$Mo($\gamma$,n) reaction is presented. Further analysis will complete the experimental database for the ($\gamma$,n) production chain of the p-isotopes of molybdenum.

\end{abstract}

\section{Motivation}
The p-nuclei between $^{74}$Se and $^{196}$Hg are produced under explosive conditions in a sequence of photodissociations of s- and r-process seeds and subsequent $\beta$-decays \cite{1992AARvLambert}. The modeling of the p-process requires a large network with more than 2,000 isotopes linked by more than 20,000 reactions. The necessary nuclear physics input includes masses of the isotopes, half-lives of the nuclei involved, and reaction rates. Only a small fraction of the required reaction rates can be determined experimentally, most of which rely on predictions from statistical model calculations \cite{2003arnould}. An experimental validation of the reaction rates included in stellar models is therefore highly desired.\\

\begin{figure}[h]
\begin{minipage}[t]{0.62\linewidth}
\includegraphics[width=1.0\linewidth]{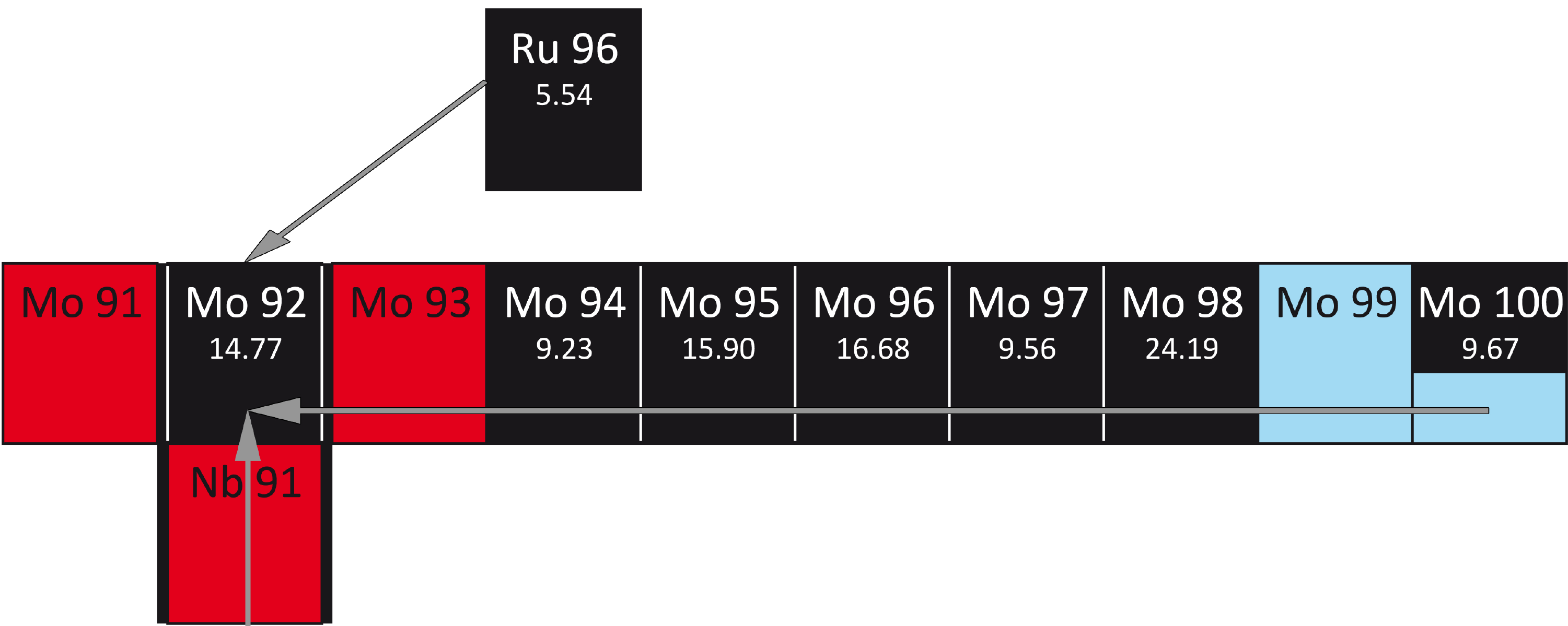}
\caption{\label{MoRegion}Portion of the reaction network leading to the production of the p-isotopes ${}^{92}$Mo and ${}^{94}$Mo.}
\end{minipage}\hspace{0.05\linewidth}%
\begin{minipage}[t]{0.33\linewidth}
\includegraphics[width=1.0\linewidth]{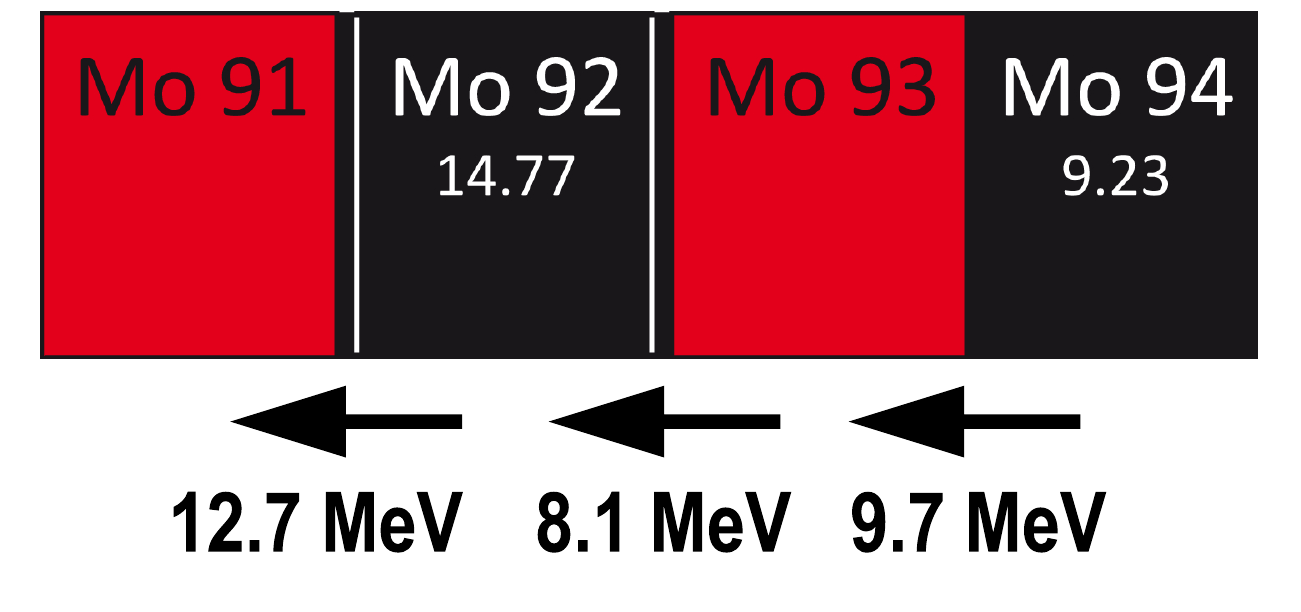}
\caption{\label{MoNeutronSepE}Neutron separation energies of the molybdenum isotopes ${}^{92}$Mo, ${}^{93}$Mo, and ${}^{94}$Mo.}
\end{minipage} 
\end{figure}

$^{92,94}$Mo and $^{96,98}$Ru are the most abundant, but not sufficiently explained p-nuclei \cite{2003arnould}. According to recent stellar model calculations, $^{94}$Mo is mainly synthesized via the ($\gamma$,n) photodisintegration chain starting from the more neutron-rich, and stable molybdenum isotopes \cite{2011ApJTravaglio} (Fig.~\ref{MoRegion}). In order to understand the abundance ratio of $^{94}$Mo to $^{92}$Mo, the determination of the cross section of the $^{94}$Mo($\gamma$,n) reaction is necessary. If this reaction is possible in a certain stellar environment, the photodisintegration of $^{93}$Mo will follow immediately due to the lower neutron separation energy (Fig.~\ref{MoNeutronSepE}). The closed neutron shell at $^{92}$Mo terminates the photoneutron chain of the molybdenum isotopes. 

The ${}^{92,93,94,100}$Mo($\gamma$,n) reactions were measured via Coulomb Dissociation at the LAND/R${}^{3}$B setup at GSI Helmholtzzentrum f\"{u}r Schwerionenforschung in Darmstadt, Germany. The ($\gamma$,n) reactions are studied in inverse kinematics. The nuclei under investigation are prepared as an ion beam, allowing radioactive nuclei to be investigated. The aim of the present experiment was the validation of the method of Coulomb Dissociation by comparing the results with data from photoactivation measurements \cite{2006EPJAkerstinsonnabend}. The ($\gamma$,n) cross section of the unstable isotope $^{93}$Mo could be determined for the first time \cite{2010NICOlgashort}.

\section{Experimental method}\label{sec::exp}
\begin{figure}[htb]
\begin{minipage}{0.50\linewidth}
\includegraphics[width=1.0\linewidth]{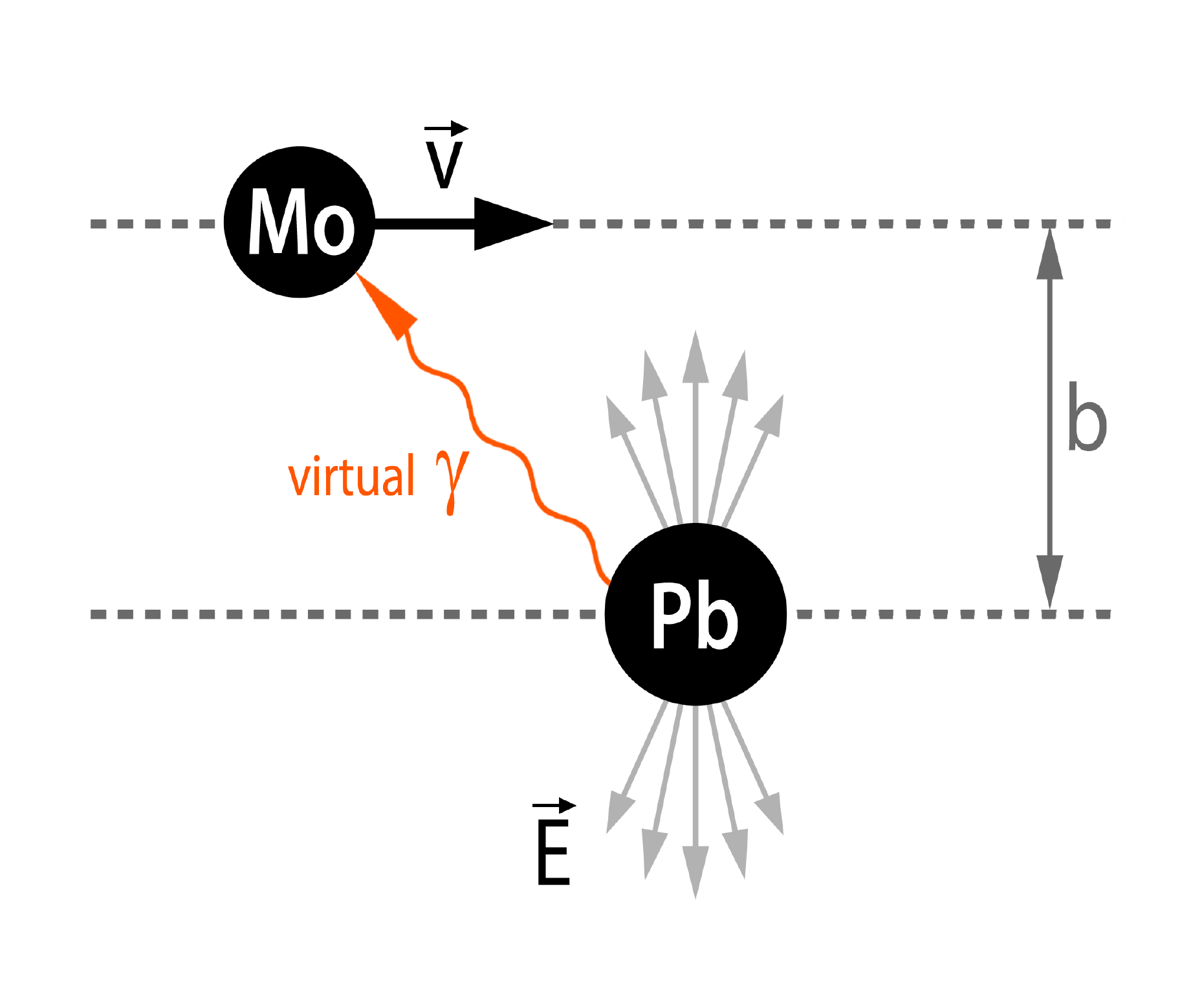}
\end{minipage}\hspace{1.cm}%
\begin{minipage}{0.44\linewidth}\caption{\label{CoulombDiss}Schematic view of the Coulomb excitation process of a molybdenum nucleus. The Mo projectile impinges on a lead target. The target's electromagnetic field $\vec{E}$ seen by the projectile is Lorentz-contracted in the direction of motion. In a peripheral collision (large impact parameter b) the projectile can be excited by a virtual photon.}
\end{minipage}
\end{figure}

Most nuclei involved in photodissociation reactions in stellar nucleosynthesis networks are unstable and cannot be prepared as a sample for experiments using real photons. One solution is to study the ($\gamma$,n) reaction in inverse kinematics: The nucleus under investigation hits a high-Z target and interacts with the time-varying Coulomb field. This interaction can be interpreted as an absorption of a virtual photon (Fig.~\ref{CoulombDiss}) \cite{BetulaniBaurCD1988}. The absorption probability can be translated into a ($\gamma$,n) cross section.\\

\begin{figure}[htb]
\begin{center}
\includegraphics[width=15cm]{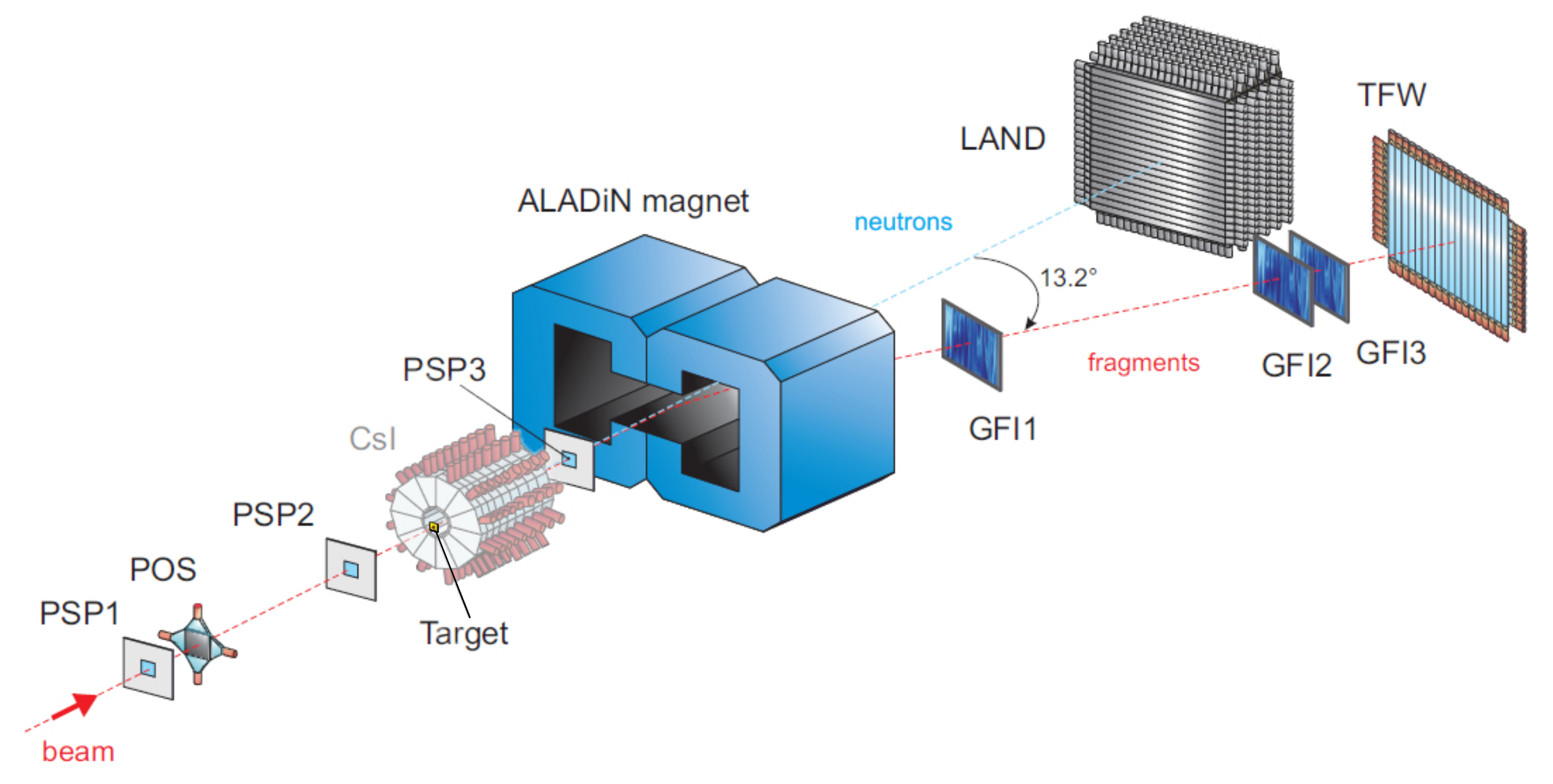}
\caption{\label{Setups295}LAND/R${}^{3}$B experimental setup for the measurement of ${}^{92,93,94,100}$Mo($\gamma$,n) by Coulomb Dissociation \cite{2012ErshovaPhD}.}
\end{center}
\end{figure}

The experimental setup is shown in Fig.~\ref{Setups295}. The ion beam enters the experimental area and is tracked in time by the POS scintillation detector. Position sensitive pin diodes (PSP)~\cite{2009NIMPSPshort} detect the position of the ion as well as its charge via energy loss in the detector. The heavy fragment is deflected by the ALADiN magnet (A Large Acceptance Dipole magNet) after the ($\gamma$,n) reaction and hits three scintillating fibre detectors (GFI)~\cite{1998NIMGFIshort} where the horizontal position is measured. The TFW (Time-of-Flight Wall) at the end of the fragment arm provides position, time and charge information. The emitted neutron remains unaffected by the magnetic field and is detected by the Large Area Neutron Detector (LAND)~\cite{1992NIMLANDshort}.

\section{Analysis and preliminary results}

Several steps are needed to extract the cross section from the data of the complex experimental setup. 

\subsection{Selection of the reaction channel}\label{sec::ReactionChannel}

The reaction channel ${}^{94}$Mo($\gamma$,n)${}^{93}$Mo has to be selected by several conditions applied to the data. A primary beam of ${}^{94}$Mo entered the experimental area. To reject breakup events in the material in front of the target, the incoming projectile ${}^{94}$Mo is identified by the charge (Z) measured by PSP1 and PSP2. In the outgoing channel, the heavy reaction fragment must be identified by its charge and mass. The fragment charge is selected by data obtained by the PSP3 and the TFW. The fragment mass is determined by tracking of the ions through the magnetic field, using position information provided by the PSP and GFI detectors. Furthermore, one neutron has to be detected by LAND.

\subsection{Background subtraction}

In addition to Coulomb excitation, the projectile may experience nuclear interactions in the target and reactions with the material outside of the target. In order to determine these background contributions, two additional measurements were performed: one without target (``empty target") and one with a carbon target. Due to the lower atomic number of carbon, the induced electromagnetic excitations are much weaker compared to lead. The cross section of the nuclear interaction is related to the size (radii) of the projectile and of the target nuclei. Therefore, the interaction of the projectile with the carbon target is considered to be purely nuclear \cite{2010BertulaniNuclAstro}. The background distributions are normalized and subtracted from the lead target distributions. The Coulomb Dissociation cross section $\sigma_{CD}$ is determined by
\begin{equation} 
	\sigma_{CD} = \left( \frac{M_{Pb}}{d_{Pb}N_{A}} \right) p_{Pb} - \left( \alpha \cdot \frac{M_{C}}{d_{C}N_{A}} \right) p_{C} - \left( \frac{M_{Pb}}{d_{Pb}N_{A}} - \alpha \cdot \frac{M_{C}}{d_{C}N_{A}} \right) p_{Em},
	\label{eq:sigmaCD}
\end{equation} 
where $p$ is the interaction probability, $M$ the molar mass of the target material [g mol${}^{-1}$], $d$ the target thickness, $N_{A}$ the Avogadro number [mol${}^{-1}$] and $\alpha$ the nuclear scaling factor. The indices refer to the different targets. In the analysis presented here, $\alpha$ is calculated according to the ``black disc model"~\cite{1993AumannPRC}, in which the interacting nuclei are considered fully opaque to each other. 
\begin{equation}
	\alpha = \frac{A^{\frac{1}{3}}_{P} + A^{\frac{1}{3}}_{Pb}}{A^{\frac{1}{3}}_{P} + A^{\frac{1}{3}}_{C}}.
\end{equation}
Here, $A_{P}$, $A_{Pb}$, and $A_{C}$ represent the mass numbers of the projectile, the lead and the carbon target, respectively. The model yields $\alpha$ = 1.53 for the ${}^{94}$Mo projectile.

\subsection{Preliminary integrated Coulomb Dissociation cross section}

For the determination of the integrated Coulomb Dissociation cross section, a distribution of any observable derived according to Eq.~\ref{eq:sigmaCD} can be used. In the preliminary analysis presented here, the neutron kinetic energy in the fragment rest frame $E_{n}$ was chosen. It is calculated on an event-by-event basis with the kinematics of the incoming ion ${}^{94}$Mo, the outgoing fragment ${}^{93}$Mo and the evaporated neutron considering the Q value of the reaction:
\begin{equation}
	E_{n} = \sqrt{m_{93Mo}^{2} + m_{n}^{2} + 2 \cdot \gamma_{93Mo} \gamma_{n} \cdot m_{93Mo} m_{n} \cdot \left( 1 - \beta_{93Mo} \beta_{n} \cos{\theta_{93Mo,n}} \right)} c^{2} - m_{94Mo} c^{2} - Q.
\end{equation}

The spectrum was corrected for the LAND efficiency and acceptance. The LAND efficiency and acceptance as a function of the neutron kinetic energy was simulated considering the nominal efficiency of LAND, which was determined in a calibration experiment, the inefficiency resulting from switched-off paddles, and the limited geometrical acceptance of the detector. The data were taken from Ref.~\cite{2012ErshovaPhD}. \\

\begin{figure}[htb]
\begin{minipage}{0.52\linewidth}
\includegraphics[width=1.0\linewidth]{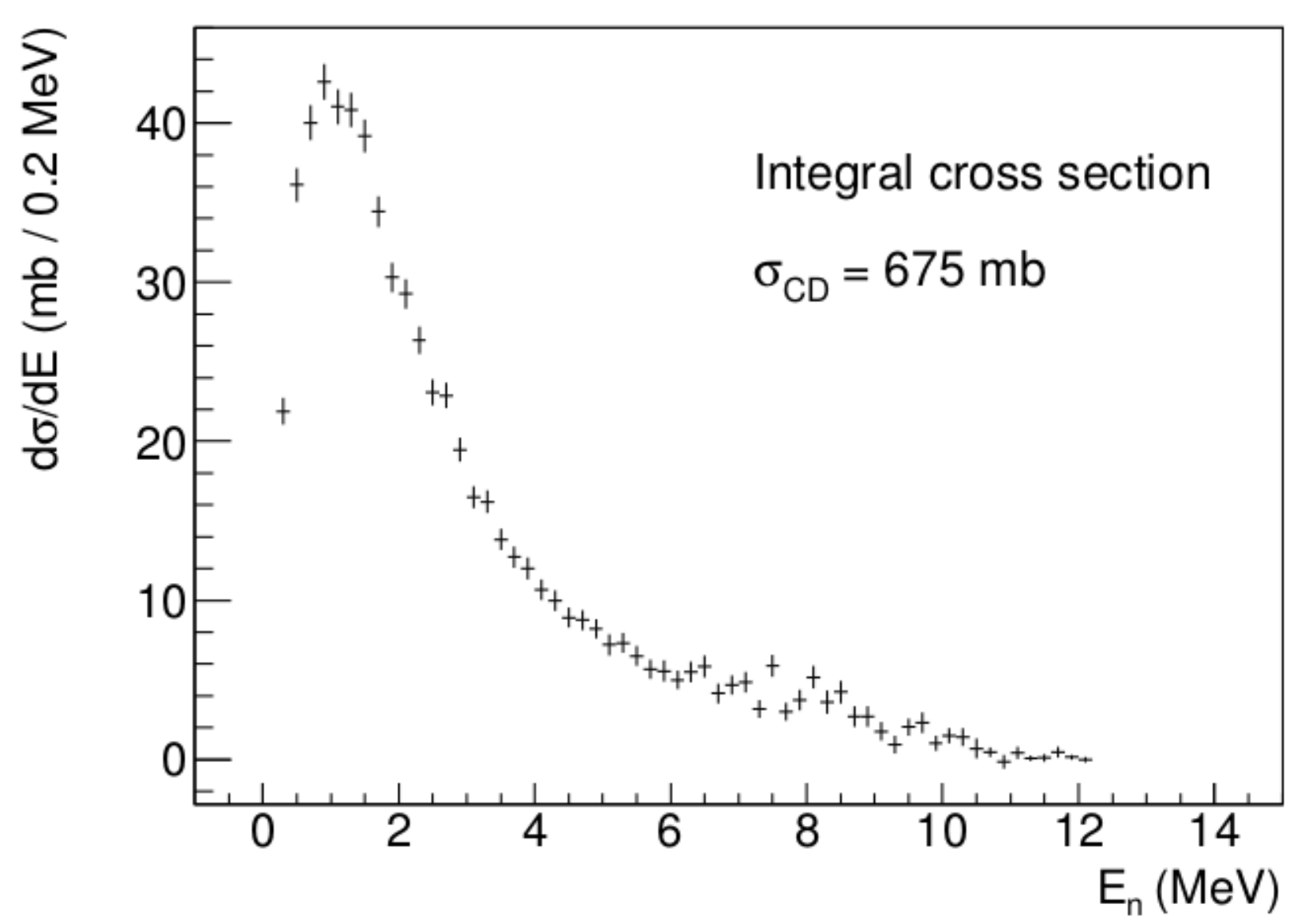}
\end{minipage}\hspace{1.cm}%
\begin{minipage}{0.42\linewidth}\caption{\label{CoulexExE}Preliminary Coulomb Dissociation cross section of the reaction Pb(${}^{94}$Mo,${}^{93}$Mo+n)Pb at a beam energy of \mbox{500 MeV}. The errors indicated are purely statistical.}
\end{minipage}
\end{figure}

The derived spectrum is shown in Fig.~\ref{CoulexExE}. Its integral delivers a preliminary integrated Coulomb Dissociation cross section of \mbox{675 mb} for the reaction Pb(${}^{94}$Mo,${}^{93}$Mo+n)Pb at a beam energy of \mbox{500 MeV}.

\section{Comparison to results from photoabsorption measurements}
\begin{figure}[htb]
\begin{minipage}{0.52\linewidth}
\includegraphics[width=1.0\linewidth]{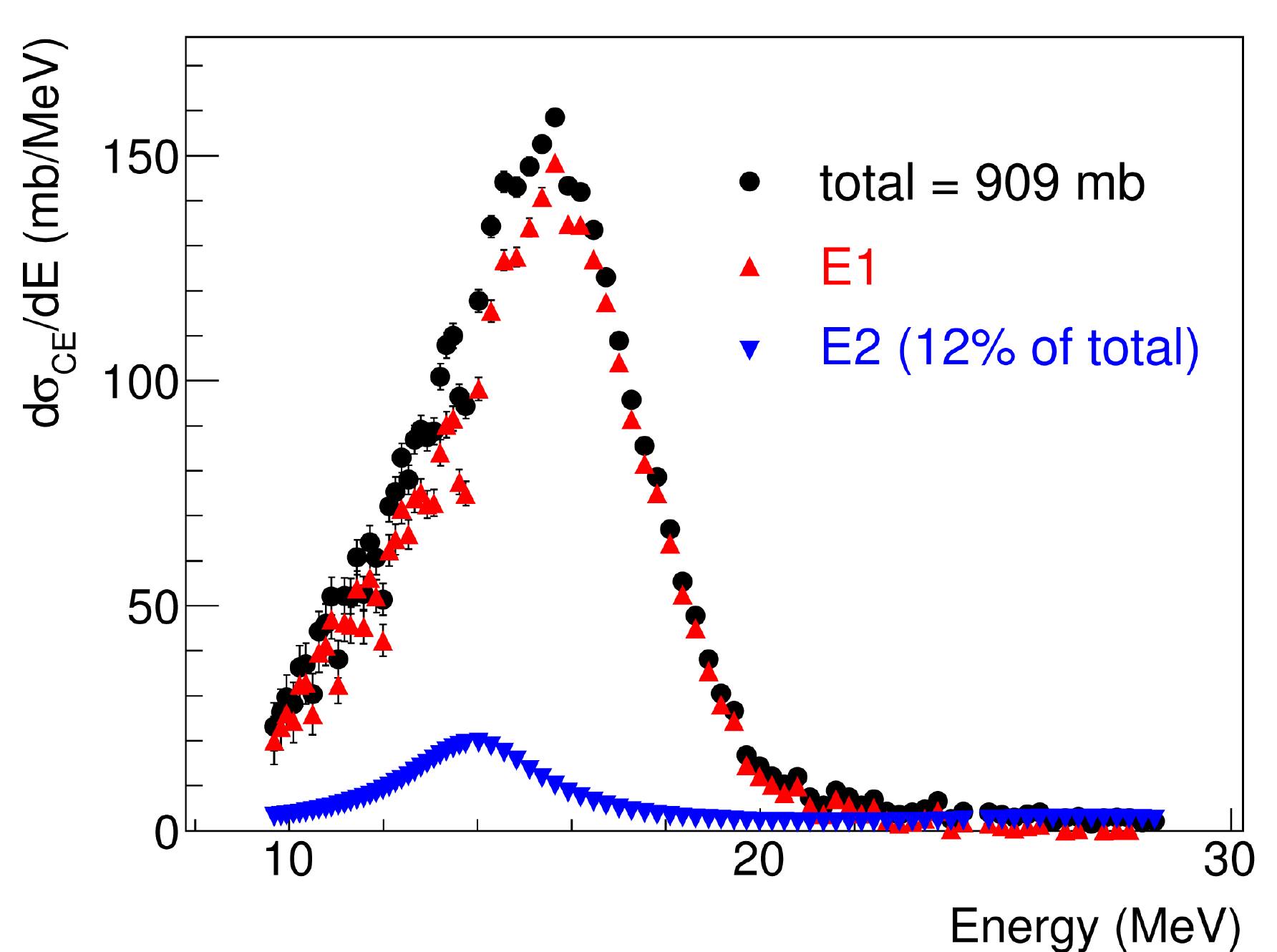}
\end{minipage}\hspace{1.cm}%
\begin{minipage}{0.42\linewidth}\caption{\label{BeilCoulex94Mo}Coulomb excitation cross section of ${}^{94}$Mo derived from the data of Beil \textit{et al.}~\cite{1974Beil}. The data were obtained by a convolution of the E1 and E2 photoabsorption cross sections with the corresponding calculated virtual photon field. The E2 component was determined from the systematics for Giant Quadrupole Resonances \cite{2012ErshovaPhD}.}
\end{minipage}
\end{figure}

The results for the reactions ${}^{92}$Mo($\gamma$,n) and ${}^{100}$Mo($\gamma$,n) are in good agreement with measurements performed with real photons~\cite{2010NICOlgashort}. According to the method described in~\cite{2012ErshovaPhD}, the photoabsorption data available for the reaction ${}^{94}$Mo($\gamma$,n)${}^{93}$Mo from Beil \textit{et al.}~\cite{1974Beil} were converted to Coulomb excitation spectra for the E1 and E2 components using the systematics for Giant Quadrupole Resonances. The data are presented in Fig.~\ref{BeilCoulex94Mo}. 

From the photoabsorption data, a Coulomb Dissociation cross section of \mbox{909 mb} is calculated. The ratio of the integral and of the integrated cross section deduced from the experimental data was found to be 0.74. As a first approximation, the obtained value agrees with the normalization factor suggested by Berman \textit{et al.}~\cite{1975Berman} and Erhard \textit{et al.}~\cite{2010Erhard}, that needs to be applied to the photoabsorption data by Beil \textit{et al.} The factor was confirmed in Ref.~\cite{2012ErshovaPhD} for ${}^{92}$Mo and ${}^{100}$Mo. The results of further analysis of the data for the ${}^{94}$Mo($\gamma$,n)${}^{93}$Mo reaction, of additional efficiency corrections and of a detailed study of the systematical errors will provide a full picture of the Coulomb Dissociation experiment of the ${}^{92,93,94,100}$Mo($\gamma$,n) reactions at the LAND/R${}^{3}$B setup.

\section{Summary and outlook}
The Coulomb Dissociation (CD) cross sections of the stable isotopes ${}^{92,94,100}$Mo and of the unstable isotope ${}^{93}$Mo were measured at the LAND/R${}^{3}$B setup at GSI Helmholtzzentrum f\"{u}r Schwerionenforschung in Darmstadt, Germany. The CD cross sections obtained for $^{92}$Mo and $^{100}$Mo and the preliminary value for $^{94}$Mo are in good agreement with experiments using real photons, thus validating the method of Coulomb Dissociation. The result of $^{93}$Mo($\gamma$,n) is particularly important since the corresponding cross section has not been measured before. The results from the ongoing analysis of $^{94}$Mo($\gamma$,n)$^{93}$Mo will complete the analysis of this series of measurements, hence completing the experimental database for the ($\gamma$,n) production chain of the p-isotopes of molybdenum.

\ack
This project was supported by the Helmholtz International Center for FAIR, the Helmholtz Young Investigator Group VH-NG-327, DFG (SO907/2-1) and HGS-HIRe.

\section*{References}
\providecommand{\newblock}{}

\end{document}